\documentclass[aps,prl,twocolumn,showpacs]{revtex4-1}
\usepackage{acronym}
\usepackage{graphicx} 
\usepackage{gensymb} 
\usepackage{textcomp}
\usepackage{color}

\begin{document}

\title{Shape control of QDs studied by cross-sectional scanning tunneling microscopy}

\author{J.G. Keizer}
\author{M. Bozkurt}
\author{J. Bocquel}
\author{P.M. Koenraad}
\affiliation{Department of Applied Physics, Eindhoven University of Technology, P.O. Box 513, NL-5600 MB, Eindhoven, The Netherlands}

\author{T. Mano}
\author{T. Noda}
\author{K. Sakoda}
\affiliation{National Institute for Materials Science, 1-2-1 Sengen, Tsukuba, Ibaraki 305-0047, Japan}

\author{E.C. Clark}
\author{M. Bichler}
\author{G. Abstreiter}
\author{J.J. Finley}
\affiliation{Walter Schottky Institut, Technische Universit\"at M\"unchen, Am Coulombwall 3, D-85748 Garching, Germany}

\author{W. Lu}
\author{T. Rohel}
\author{H. Folliot}
\author{N. Bertru}
\affiliation{INSA, Universit\'{e} Europ\'{e}enne de Bretagne, 20 Avenue des Buttes de Co\"esmes, F-35043 Rennes Cedex, France}

\begin{abstract}
In this cross-sectional scanning tunneling microscopy study we investigated various techniques to control the shape of self-assembled quantum dots (QDs) and wetting layers (WLs). The result shows that application of an indium flush during the growth of strained InGaAs/GaAs QD layers results in flattened QDs and a reduced WL. The height of the QDs and WLs could be controlled by varying the thickness of the first capping layer. Concerning the technique of antimony capping we show that the surfactant properties of Sb result in the preservation of the shape of strained InAs/InP QDs during overgrowth. This could be achieved by both a growth interrupt under Sb flux and capping with a thin GaAsSb layer prior to overgrowth of the uncapped QDs. The technique of droplet epitaxy was investigated by a structural analysis of strain free GaAs/AlGaAs QDs. We show that the QDs have a Gaussian shape, that the WL is less than 1\,bilayer thick, and that minor intermixing of Al with the QDs takes place.       
\end{abstract}

\maketitle

\section{Introduction}

In the last decade the fabrication of self-assembled quantum dots (QDs) has been intensively studied. The interest has been, and still is, stimulated by applications of QDs in optoelectronic devices. From previous studies it is well known that the optical and electronic properties of QDs are strongly affected by their size, shape, and material composition. Despite years of intense studies, control over these properties remains difficult. One major problem is the change in QD morphology during the growth of the capping layer. Traditionally control over the QD height, one aspect of the change in morphology, can be achieved with monolayer precision by the double capping method \cite{Paranthoen2001} or the so-called indium flush method \cite{Wasilewski1999}, a variation on the former technique. In the latter technique the growth of the capping layer is interrupted, at which point the temperature is raised to remove any surface resident indium. This effectively locks the height of the QD and prevents any further In segregation \cite{Keizer2010}. Another approach in shape control of QDs is the use of surfactants. Recently, antimony has received a great deal of attention in its role during the capping process due to its surfactant properties. It has been shown that Sb reduces the surface diffusion of other atoms but without getting incorporated itself \cite{Harmand2004}, allowing the achievement of fully pyramidal shaped QDs \cite{Ulloa2007a}. Yet another approach to gain control over the erosion of QDs during overgrowth and thus over the shape of the QD is the removal of the driving force: lattice strain. This can be achieved in lattice matched QDs grown by droplet epitaxy. First reported by Koguchi \textit{et al.} \cite{Koguchi1991}, this technique involves low temperature growth of unstrained group III-element droplets that are subsequently crystallized into QDs by incorporation of group V-elements. It has been shown that this technique can be used to grow nearly pure nanostructures \cite{Keizer2010a} with a typical size distribution of 10--20\% \cite{Mano2005}.

In this paper the techniques of indium flush, antimony capping, and droplet epitaxy are studied by means of cross-sectional scanning tunneling microscopy (X-STM). We first investigate the degree of control that can be achieved over the height of InGaAs/GaAs QDs and the wetting layer (WL) by means of an indium flush. We then go on by showing that antimony capping can be employed to prevent QD erosion during the capping process of InAs/InP QDs. Finally, the intermixing in, and the shape of, GaAs/AlGaAs QDs grown by droplet epitaxy is examined in detail.

\section{Experimental setup}

All X-STM measurements were performed at room temperature under UHV ($p<6\times10^{-11}$\,mbar) conditions with an Omicron STM-1, TS2 Scanner. The STM was operated in constant current mode on {\it in situ} cleaved (110)-surfaces. Electrochemically etched tungsten tips were used. The QD layers were grown by molecular beam epitaxy (MBE). The details of the growth procedure for the different material systems will be described separately in their corresponding sections.

\section{Indium flush} 

The material system used to investigate the indium flush technique consists of InGaAs QD layers grown by MBE on an \textit{n}-type GaAs (001) orientated substrate. An undoped GaAs buffer layer of 420\,nm was grown at 690\,$^{\circ}$C, followed by a growth interruption of approximately 2\,min that allowed the temperature to be lowered to 600\,$^{\circ}$C, the nominal growth temperature of the QD layers. Following this, three sequences consisting of four QD layers of 1.98\,nm (7\,ML) In$_{0.5}$Ga$_{0.5}$As were deposited. During the whole growth process the As flux was kept constant at a pressure of $1.26\times10^{-5}$\,mbar. Three out of the four QD layers were grown with the indium flush method which consists of the following procedure. First the QD layers are partially capped with a GaAs layer of which the thickness was varied. Next, the temperature is raised to 650\,$^{\circ}$C for 30\,s and lowered again to the nominal growth temperature after which a second GaAs capping layer is deposited. In total, the annealing step takes place over a time window of $\approx180$\,s. The total structure was capped with 200\,nm GaAs.

\begin{figure}[b]
\includegraphics[width=8.6cm]{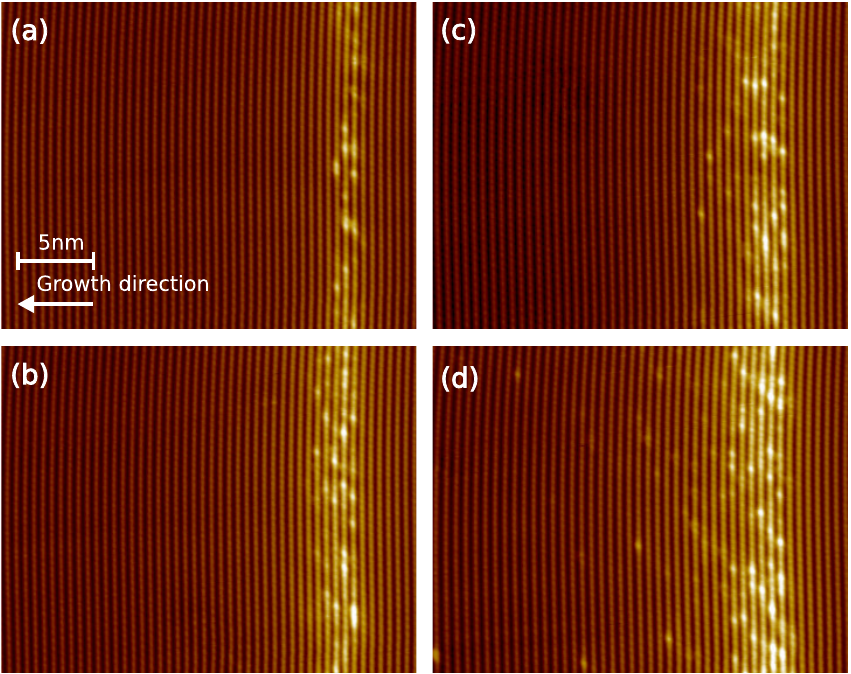}
\caption{\label{figure_1}X-STM images of the InGaAs WL as a function of the capping layer thickness. (a) 2\,nm (b) 3\,nm (c) 6\,nm first capping layer thickness and (d) conventionally grown capping layer.}
\end{figure}

\begin{figure}
\includegraphics[width=8.6cm]{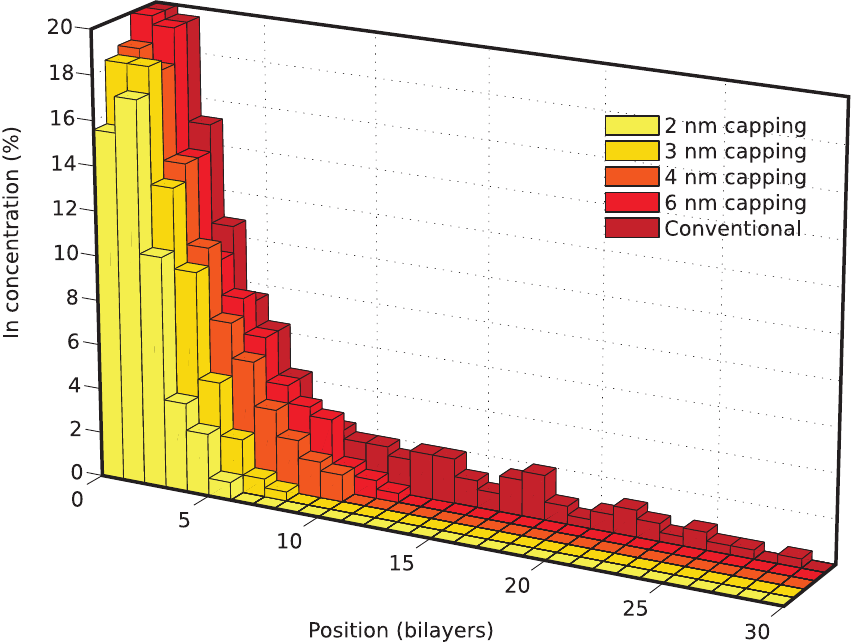}
\caption{\label{figure_2}In segregation as a function of first capping layer thickness and bilayer position from the start of the WL.}
\end{figure}

\begin{figure}[t]
\includegraphics[width=8.6cm]{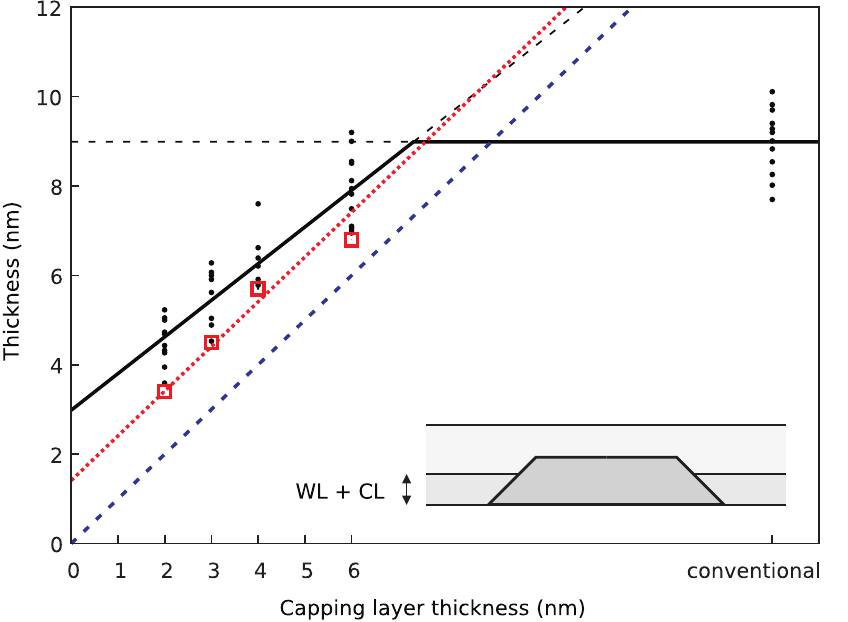}
\caption{\label{figure_4}QD height (black points) as a function of the thickness of the first capping layer. The black line is a linear fit. The dotted red line represents the sum of the critical layer thickness (5\,ML) and the first capping layer thickness (dashed blue line). The experimentally determined average thickness of the final WL is given by the open red boxes.} 
\end{figure}

We begin by analysing the WL thickness and composition. In figure~\ref{figure_1}, four typical X-STM images of WLs grown with different capping layer thicknesses are depicted. Even without any statistical analysis it is evident that the height of the WL can be controlled by varying the height of the capping layer. In addition, the In segregation appears to terminate abruptly in case of WLs that underwent an indium flush. This is a clear indication that most of the surface resident In is removed during the flush step, preventing further segregation. In order to make our analysis more quantitative we counted and marked the bilayer position from the start of the WL for approximately 3000 In atoms. An $\approx400$\,nm cross-sectional region of each WL present in the sample was analysed in this manner. In figure~\ref{figure_2}, the result of our statistical analysis is shown. The conventionally grown WL exhibits the expected exponential decay of the In concentration and In segregation length ($\approx25$\,nm) \cite{Offermans2005b}. In contrast, the WLs grown with the indium flush procedure show a stronger decay and shorter segregation length. This implies that In segregates out of the WLs and leaves the surface during the indium flush step. This additional loss of already buried In is strongest in case of the thinnest capping layer. The total amount of In that remains after flush-off is thus strongly dependent on the capping layer thickness due to desorption and additional segregation. The thickness of the final WL is found to be 6, 8, 10, 12 bilayers for 2, 3, 4, 6\,nm thick first capping layers, respectively. Note, that the statistical analysis presented in figure~\ref{figure_2} reveals that the WL extends further than is expected from figure~\ref{figure_1}. It is reported that the critical WL thickness for In$_{0.5}$Ga$_{0.5}$As QD formation is $\approx5$\,ML (1.4\,nm) \cite{Snyder1992}. We assume that all the In that is deposited after reaching this critical thickness goes into the formation of QDs. If we add the thickness of the critical layer ($d_{\mathrm{crit}}$) to the thickness of the first capping layer ($d_{\mathrm{cap}}$) and compare the resulting sum with the experimentally found thickness of the final WL we find good agreement. This is depicted quantitatively by the dotted red line and open red boxes in figure~\ref{figure_4}, that show $d_{\mathrm{crit}}+d_{\mathrm{cap}}$ and the experimentally determined average WL thickness, respectively. This result shows that In segregation beyond the position of the flush is completely suppressed; In is absent in the final GaAs capping layer.

\begin{figure}[t]
\includegraphics[width=8.6cm]{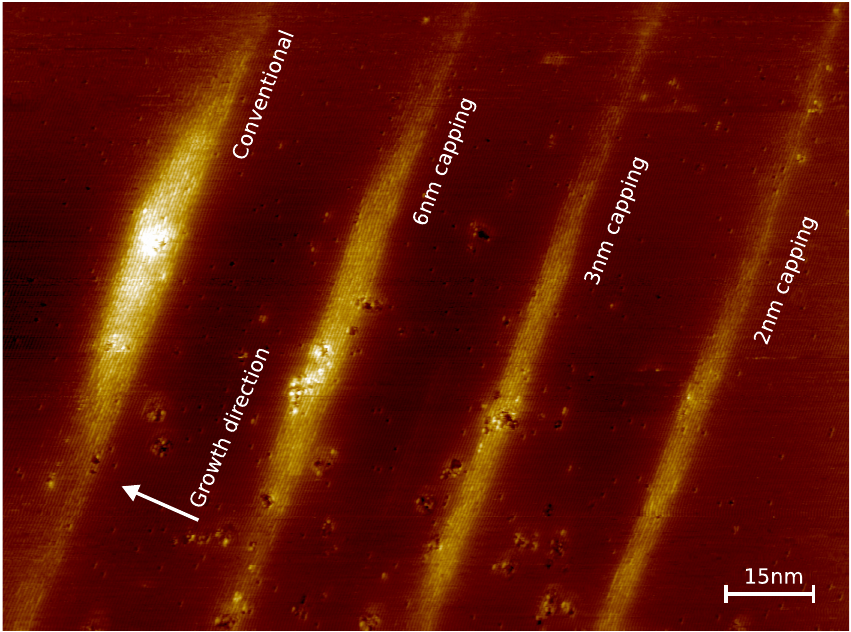}
\caption{\label{figure_3}X-STM image of one conventionally grown QD and three QDs grown with an indium flush step incorporated in the growth process. The thickness of the first capping layer was varied.}
\end{figure}

In order to determine the influence of the indium flush step on the structural properties of the QDs we determined the width and height of a total of 48 cleaved QDs. The width of the QDs ranged up to 100\,nm. The height of the conventionally grown QDs was found to vary between 7 and 10\,nm. The QD layers were found to be weakly coupled, as one would expect with the GaAs spacer layer being 30\,nm thick and slightly strained InGaAs QDs \cite{Bruls2003}, resulting in occasional stacking of the QDs. Figure~\ref{figure_3} shows one of the sequences consisting of four QD layers where the QDs are stacked. The thickness of the first capping layer was varied in the first three QD layers from 2, 3 to 6\,nm. The last layer is a conventionally grown QD layer, i.e. without the application of an indium flush step. As can be seen, the application of an indium flush step results in lowering of the QD height as compared to the conventionally grown QDs. The shape of the conventionally grown QD is lens like as expected for typical InGaAs QDs \cite{Wang2004}. The heights of all the observed QDs as a function of the first capping layer thickness are plotted in figure~\ref{figure_4}. Since, the lateral width of all the observed QDs was found to be of the order of 60\,nm, we can assume that none of the QDs is cleaved through their edge and that figure~\ref{figure_4} represents the spread in the height distribution of the QDs due to the growth process. We found a linear relation between the QD height and the first capping layer thickness up to $\approx 7$\,nm, indicated by the black line. Since increasing the first capping layer beyond a height of 7\,nm would make the growth procedure resemble conventional growth, we expect the QD height to saturate at this value. This is indicated in figure~\ref{figure_4} by the black horizontal line which represents the average height of the conventionally grown QDs. Note, that the average QD height in the absence of the first capping layer intersects at an offset. Moreover, the QDs are found to be higher than the final WL, see the dashed red line, open red boxes, and inset of figure~\ref{figure_4}. From this we conclude that the performed indium flush is incomplete.

\section{Antimony capping}

In the previous section we have shown that the indium flush technique can be used to lower the height of InGaAs QDs. We continue with an investigation of antimony capping, a technique that can be employed to prevent QD erosion during capping. Four InAs QD layers separated by 30 nm of InP were grown on an \textit{n}-type (311)B oriented InP substrate by solid source MBE. The growth temperature was set at 450$^\circ$C. The QDs were formed by the deposition of 2.1\,ML (001) equivalent monolayers. After QD formation, a 30\,s growth interrupt (GI) under As pressure was performed for all layers. Previously, it has been shown that As/P exchange is limited under such GI conditions \cite{ulloa2007c}. The first QD layer was overgrown with an InP capping layer. This first QD layer will be considered as the reference layer. For the second QD layer a growth interrupt under a Sb beam equivalent pressure of 2.7\,$\times\,10^{-7}$\,Torr (GISb) was performed during 30\,s before the growth of the InP capping layer. For the third and fourth layers, respectively a 1 nm and 2 nm GaAs$_{0.51}$Sb$_{0.49}$ (lattice matched to InP) thick layer was deposited after a 5\,s GISb.

\begin{figure}
\includegraphics[width=8.6cm]{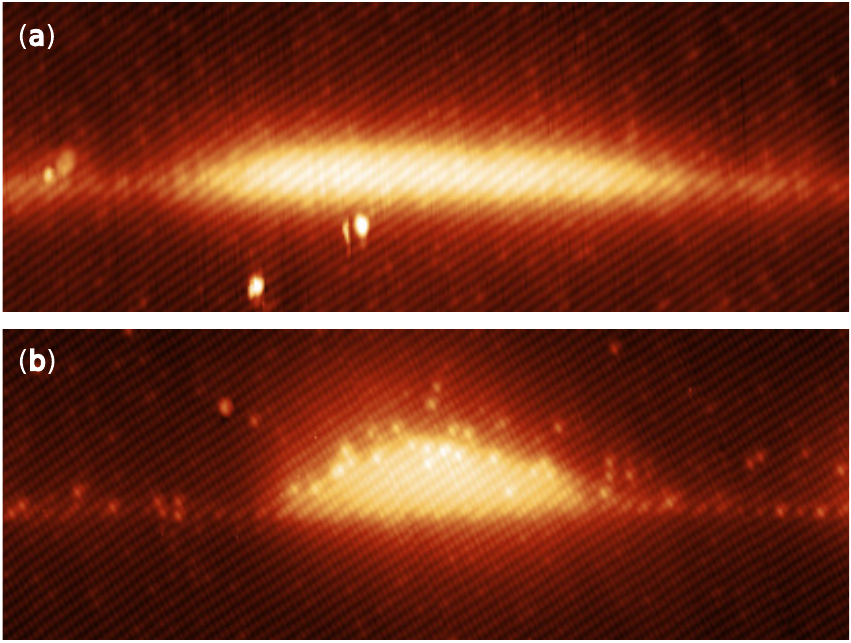}
\caption{\label{figure_5}Two 60\,nm$\,\times$\,15\,nm X-STM images. (a) InAs QD capped with InP  after a 30\,s GI. (b) InAs QD capped with InP after a 30\,s GI $+$ 30\,s GISb. The bright spots correspond to Sb atoms.}
\end{figure}

\begin{figure}
\includegraphics[width=8.6cm]{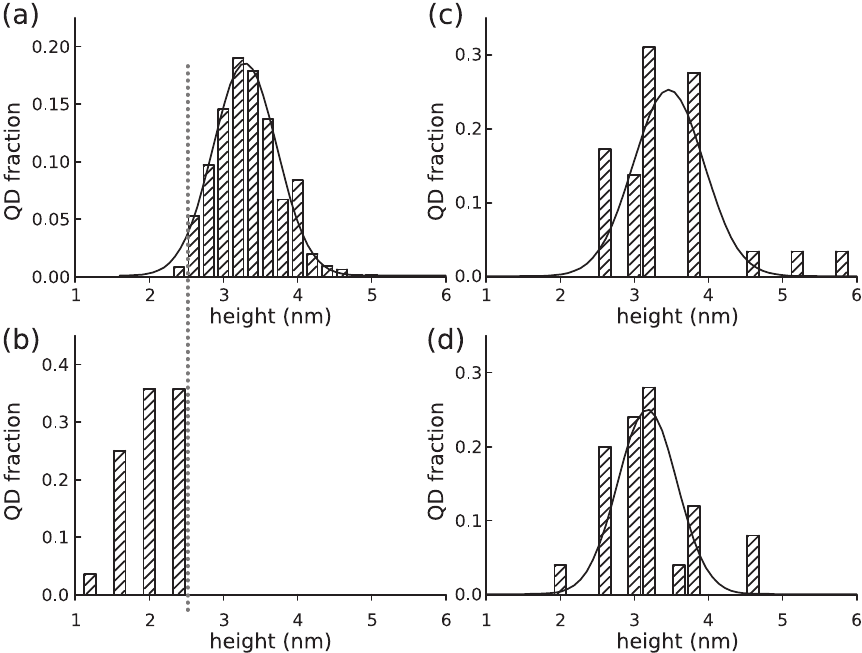}
\caption{\label{figure_6}QD height distribution of a) uncapped InAs QDs, (b) InP capped InAs QDs, (c) InP capped InAs QDs after 30\,s GISb, and d) GaAsSb capped InAs QDs after 5\,s GISb.}
\end{figure}

In figure~\ref{figure_5}a, an X-STM image of a typical QD in the reference layer is shown. These QDs are found to have a flat top facet. The homogeneity of the contrast within the QD indicates that it consist of almost pure InAs. No digging in of the WL in the underlying material as in \cite{ulloa2007c} was observed. The intermixing at the corners is minimal, like in the case of InAs QDs in AlAs \cite{Offermans2005}. The average height and width estimated from 20 individually observed QDs in the reference layer are found to be 2.0 nm and 25\,nm, respectively. Before capping, the QDs have an asymmetric pyramidal shape, bounded by low-index facets \{001\}, \{111\}B, and \{110\} \cite{Lacombe1999}. Height histograms of the uncapped QDs deduced from AFM analysis, and height histograms of the capped QDs in the reference layer as observed by X-STM are shown in figure~\ref{figure_6}a-b. For the uncapped QDs, a Gaussian distribution centered around 3.3 nm is found, whereas after InP capping the height distribution is truncated at 2.4 nm. As demonstrated previously \cite{ulloa2007c}, the truncated distribution and the flat top facet of InAs/InP QDs are to a large extend the consequence of QD decomposition. This decomposition is driven by the strain mismatch between the InP capping layer and the InAs QDs.

Figure~\ref{figure_5}b, shows InAs QDs for which a 30\,s GISb has been performed before the InP capping layer was grown. The bright spots correspond to Sb atoms remaining in the InP capping layer and in the InAs QDs after the GISb and the succeeding growth of the capping layer. Given the total amount of Sb supplied to the surface and the observed amount of Sb after capping, we conclude that a large part is desorbed during overgrowth. Segregation of the small fraction of Sb that gets incorporated in the InP capping layer is clearly shown. Within the QDs the back diffusion of Sb is negligible and a preferential incorporation of Sb is observed at the outermost layers of the QDs. Again, the InAs QD corners appear well defined with minimal intermixing and formation of an InAsP alloy, just as is the case with the QDs in the reference layer. The presence of Sb on the surface induces changes on the QD shape; the mean height is now 3.5\,nm (see figure~\ref{figure_6}c) and the mean diameter 21\,nm, corresponding to the dimensions of the uncapped QDs. We can explain the observed shape preservation by the well documented surfactant effect of Sb atoms \cite{Aivaliotis2007,Harmand2004}. An Sb surfactant can limit the in-plane diffusion of atoms on the surface. Accordingly, the InAs diffusion from the QD apex to the periphery should be reduced due to the presence of Sb atoms on the surface. This freezing of the mass transport on the growth front results in the preservation of the shape of the uncapped QDs.

\begin{figure}
\includegraphics[width=8.6cm]{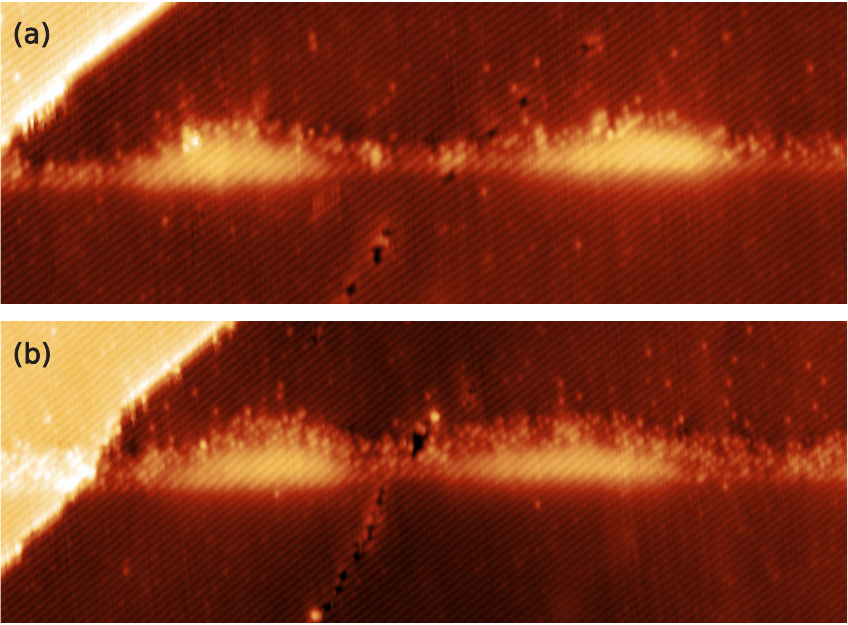}
\caption{\label{figure_7}Two 100\,nm$\,\times$\,20\,nm X-STM images of InAs QDs and the WL capped with (a) 1\,nm and (b) 2\,nm GaAsSb after a 30\,s GI $+$ 5\,s GISb. A single step edge is visible at the left side in both images.}
\end{figure}

X-STM images of the third and fourth QD layers are shown in figure~\ref{figure_7}a-b. These layers were, after 30\,s GI $+$ 5\,s GISb, capped with a thin layer of GaAsSb (lattice matched to InP). As was the case with the 30\,s GISb, the QDs in these layers are taller than those in the reference layer; for both layers, an average height of 3.2 nm (see figure~\ref{figure_6})d and a base diameter of 21 nm are deduced, corresponding to the dimensions of the uncapped QDs. Again, the intermixing in the QDs is negligible. Similar shape conservation has been reported when InGaAs or GaAsSb strained capping layers are grown on InAs/GaAs QDs \cite{ulloa2007c}. In that case a phase separation is observed in the ternary capping layer on top of the QDs. In our case, the GaAsSb layer is lattice matched to the InP substrate and the observed conformal growth of GaAsSb on InAs/InP QD might be related as previously to a low group III atoms migration when Sb atoms are present on surface.

\section{Droplet Epitaxy}

Having shown that the indium flush technique and the surfactant properties of Sb allow control over the shape and height of SK-grown QDs, we now turn our attention to QDs in a lattice matched materials system. More specifically, a GaAs/AlGaAs QD layer grown on an \textit{n}-type (001) oriented GaAs substrate by droplet epitaxy. The sample was grown in the following manner. First an AlGaAs buffer layer is grown at 580\,$^{\circ}$C. Next, the sample is cooled down to 200\,$^{\circ}$C, the As flux switched off, and the As evacuated from the growth chamber. The result is an As-stabilized \textit{c}(4$\times$4) surface. Subsequently, 3.75\,ML Ga, of which the first 1.75\,ML changes the excess As into a two-dimensional GaAs layer \cite{Sanguinetti2003}, is deposited at a rate of 0.5\,ML/s. The remainder of 2\,ML will form liquid Ga droplets on the surface. Next, these droplets are crystallized into a GaAs QDs by supply of an As$_4$ flux ($2\,\times\,10^{-4}$\,Torr beam equivalent pressure). Still under As$_4$ flux, the sample is then annealed at 350\,$^{\circ}$C for 10 minutes. Subsequently the structures are capped with 50\,nm AlGaAs deposited at 350\,$^{\circ}$C, followed by a second annealing step at 650,$^{\circ}$C under As$_4$ flux for 5 minutes. This last anneal step is inserted into the growth procedure to ensure that the next layer is grown on a defect free surface. Next, another capping layer of 40\,nm is grown at 580\,$^{\circ}$C. The total structure was capped with 600\,nm GaAs. A post growth anneal step, which is usually performed to improve the optical properties of the QDs was not performed on this sample. All the images presented in this section are recorded with high negative bias ($\approx\,$-3.2\,V) between sample and STM tip. At these tunneling conditions and with the color scaling used, dark regions represent AlAs rich regions while bright regions represent GaAs rich regions.

\begin{figure}[t]
\includegraphics[width=8.6cm]{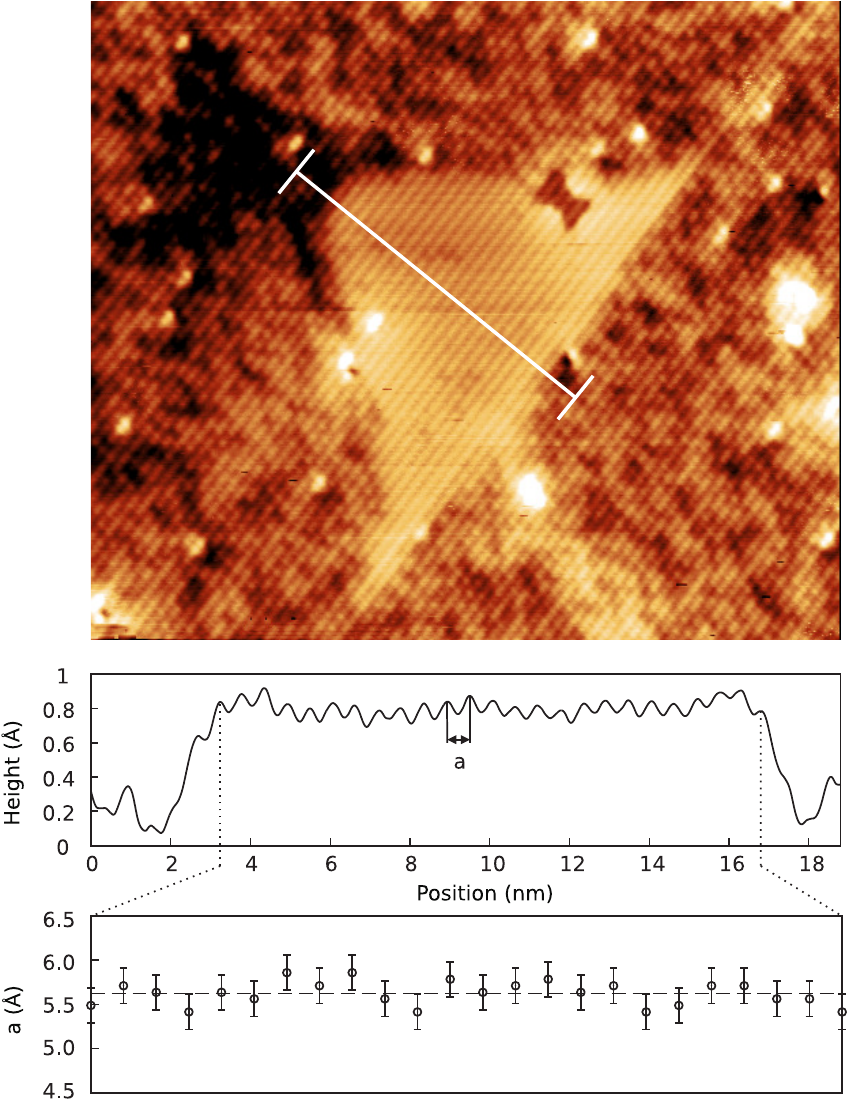}
\caption{\label{figure_8}40\,nm\,$\times$\,34\,nm topographic image of a typical GaAs/AlGaAs QD (top) and an average cross-sectional profile (top graph) and separation between bilayers (bottom graph) along the line in the top figure.}
\end{figure}

A total of 11 QDs where observed by X-STM. A typical QD is shown in figure~\ref{figure_8}. As can be seen in this topographic image, the QDs are sharply defined by abrupt interfaces. The thickness of the WL was found to be less than 1\,bilayer~\cite{Keizer2010a}, as expected. The bow tie feature is most likely a foreign atom and is of no interest in the current study. Since AlAs and GaAs are lattice matched materials, the QDs are expected to be strain free. This is checked by taking a cross-sectional profile of the QD in figure~\ref{figure_8}. Three distinct regions can be observed. From left to right: an AlAs rich region, the GaAs QD, and the AlGaAs matrix. The height difference between these regions is due to electronic contrast. More importantly, all the regions are flat, there is no outward relaxation as observed in QDs grown with lattice-mismatched systems \cite{Offermans2005b}. To further illustrate that the GaAs QD is strain free, the distance between adjacent bilayers along the cross-sectional profile was measured. For this analysis the STM piezo elements were calibrated by performing a 2D FFT on the AlGaAs matrix. The result is shown in the bottom graph of figure~\ref{figure_8}. As can be seen, there is little deviation from the expected value of 0.565\,nm (dashed line), indicating that the QD is indeed strain free. Note that there is an Al rich region on top of the QD. This can be explained by the difference in mobility of Al and Ga atoms; the Ga atoms are more mobile and will migrate along the side of the QD during capping while the Al atoms, which are less mobile, are more likely to remain on top of the QD. The driving force behind the migration of the incoming adatoms away from the top of the QD is the convex curvature of the growth front at the position of the QDs \cite{Xie1994}. Note that this different from the SK-grown QDs of the previous sections were strain induced by the lattice mismatch is the driving process.

\begin{figure}[t]
\includegraphics[width=8.6cm]{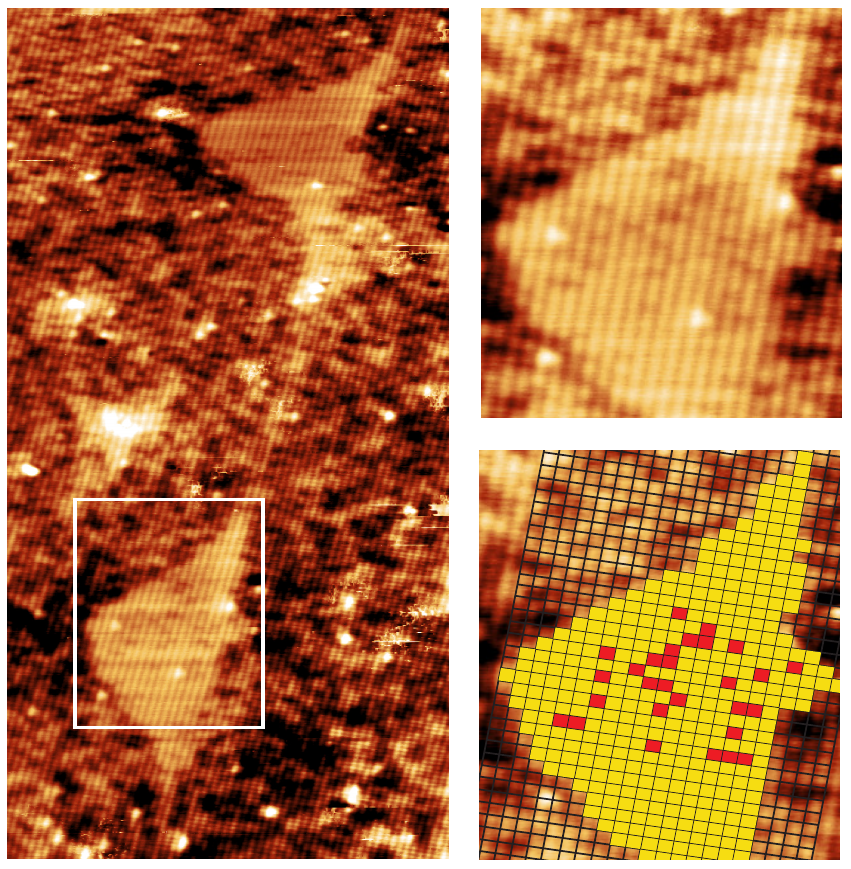}
\caption{\label{figure_9}30\,nm\,$\times$\,60\,nm topographic image (left) of two QDs. An atomic grid is overlain on top of a close up of the QD dot (right). Al and Ga atoms in the QD are indicate by respectively red and yellow squares.}
\end{figure}

Whether intermixing of Al is a factor of importance in the formation of GaAs/AlGaAs QDs grown by droplet epitaxy is a question frequently raised in the literature \cite{Sanguinetti2002,Mantovani2004,Heyn2007}. In all QDs imaged we have observed some degree of intermixing. In figure~\ref{figure_9} left panel, two typical QDs are shown. Even without further analysis it is evident that some intermixing of Al has taken place, see dark spots inside the QDs. To make a more quantitative analysis we overlaid a grid with atomic dimensions on top of a close up of the QD that showed the strongest intermixing, see figure~\ref{figure_9} right panels. On this grid, the positions of the Al and Ga atoms are marked with respectively red and yellow squares. The concentration of Al in this particular QD is determined to be 6\%. Here we would like to point out that the observed Al intermixing varied strongly from dot to dot, see for example the QD depicted in figure~\ref{figure_8} were the degree of intermixing is considerably lower, and that the 6\% can be considered as an upper limit of intermixing in these QDs.

\begin{figure}
\includegraphics[width=8.6cm]{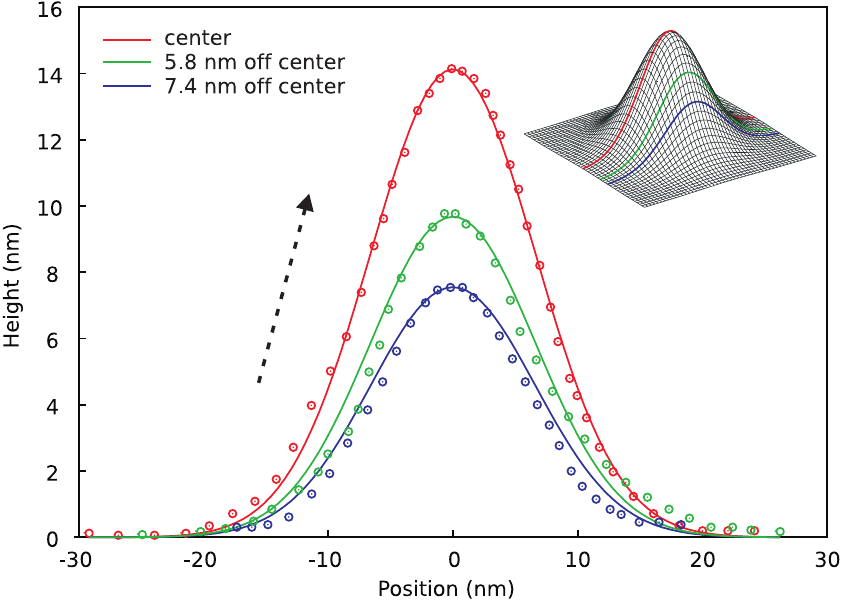}
\caption{\label{figure_10}Profile of three QDs extracted from the X-STM data (open circles). A Gaussian function is fitted to the largest QD (red line). The other two QDs (green and blue line) are assumed to have the same 3D-structure as the largest QD but cleaved off center. The projection of the (111)-direction on the cleavage plane is given by the dashed black line.}
\end{figure}

Concerning the shape of the QDs, we notice that the side facets of the observed QDs are not exactly straight. The maximum side facet angles were found to be in the range 34--55$^\circ$ per QD, were the upper limit corresponds to a \{111\} facet (54.7$^\circ$). If we assume that (1) all the QDs are approximately of equal height and (2) the observed height difference is due to the position of the cleavage plane relative to the center of the QD, this result excludes QD shapes with constant facet angles like rectangular (truncated) pyramids \cite{Bruls2002}. Since it has been reported that uncapped AlGaAs/GaAs QDs have \{111\} facets \cite{Lee1998,Mano2008}, we conclude that the shape of the QDs is somewhat changed during capping. Figure~\ref{figure_8} shows the highest QD we found. Since it is the highest, we assume that this QD is cleaved directly through its center. Consequently, we used the profile of this QD to generate a 3D-profile by fitting a Gaussian function, see figure~\ref{figure_10} (red line), and rotating it around the symmetry axis along the growth direction. Next, we checked whether the profile of all other observed QDs (illustrated for two exemplary QDs by the green and blue lines) correspond to profiles obtained by cleaving the obtained 3D-profile at specific distances from the center. As can be seen in figure~\ref{figure_10}, this is the case. From this we conclude that the observed QDs are Gaussian shaped QDs of approximately the same height but cleaved at different position from their center.

\section{Conclusion} 
To summarize, we have investigated three techniques that can be used to gain control over the shape of QDs. The indium flush technique allows control over the height of the WL and the InGaAs QDs. The resulting QDs have a flatted top facet. We have shown that not only surface resident In but also buried In that segregates out of the WL is desorbed during the indium flush. Concerning the technique of antimony capping, we have shown that a growth interrupt under Sb flux prior to capping preserves the shape of the uncapped QDs. The same could be achieved by the growth of a GaAsSb capping layer. This capping layer was found to conformally cover the growth front. In both case the preservation of QD shape is attributed to the surfactant properties of Sb. In QD layers grown by droplet epitaxy the WL was found to be less than 1\,bilayer thick. As expected in lattice-matched systems, we found no strain present in the GaAs/AlGaAs QDs. Without strain there is no driving force for QD erosion during overgrowth, resulting in high QD of which the shape was found to be Gaussian. We conclude that indium flush, antimony capping, and droplet epitaxy can all be used as a tool to both shape QDs and WL.

\end{document}